\documentclass[prb,aps,twocolumn,showpacs,amsmath,amssymb]{revtex4}
\usepackage{graphicx}
\usepackage{dcolumn}

\begin{document}

\title{Resonant Tunneling in a Dissipative Environment}

\author{Yu. Bomze, H. Mebrahtu, I. Borzenets, A. Makarovski, and G. Finkelstein}

\affiliation{Department of Physics, Duke University, Durham, NC 27708}

\begin{abstract}
We measure tunneling through a single quantum level in a carbon nanotube quantum dot connected to resistive metal leads. For the electrons tunneling to/from the nanotube, the leads serve as a dissipative environment, which suppresses the tunneling rate. In the regime of sequential tunneling, the height of the single-electron conductance peaks increases as the temperature is lowered, although it scales more weekly than the conventional $\propto T^{-1}$. In the resonant tunneling regime (temperature smaller than the level width), the peak width approaches saturation, while the peak height starts to \emph{decrease}. Overall, the peak height shows a \emph{non-monotonic} temperature dependence. We associate this unusual behavior with the transition from the sequential to the resonant tunneling through a single quantum level in a dissipative environment.

\end{abstract}

\pacs{73.23.Hk, 73.23.-b}

\maketitle

The notion of dissipation in quantum systems is important for understanding such foundations of quantum mechanics as quantum measurements and decoherence. Of particular interest is the physics of tunneling in a dissipative environment \cite{LeggetRevModPhys}, which is manifested in electronic transport phenomena as the ``environmental Coulomb blockade'' \cite{NI}. In this effect, conductance of a single nanoscale tunneling junction is suppressed at low temperatures and low bias voltage \cite{ECB_experiement,note}. In order to observe the suppression, the two leads connecting the junction to the measurement system should have the resistance comparable to the quantum unit of resistance, $h/e^2$. The tunneling electrons couple to the electromagnetic modes in the leads (referred hereafter as ``environment''), which reduces the tunneling probability. At small bias $V$ and at low temperature $T$ the differential conductance is suppressed as $G\propto \max(k_B T, eV)^{\alpha}$, where the exponent $\alpha = 2 e^2 R_e/ h$ and $R_e$ is the total resistance of the leads \cite{NI}.

Suppression of tunneling into \emph{single-wall} nanotubes has been attributed to the Luttinger liquid behavior \cite{singlewalltunnel}, while a similar effect in \emph{multi-wall} nanotubes has been discussed in terms of environmental Coulomb blockade due to the resistive nature of the nanotubes \cite{multiwalltunnel}. Our set-up is different from either one: suppression of tunneling develops in our case not due to the properties of the nanotube, but due to the high resistance of the long metal leads that connect the nanotube to the measurement set-up. The nanotube itself serves either as a tunable tunneling junction between the two leads, or as a quantum dot, depending on the gate voltage. The ability to gate the nanotube allows us to first characterize its environment and then to study the  physics of tunneling through a single quantized level in the known dissipative environment. We are particularly interested in the novel regime of resonant tunneling with dissipation. Our results confirm the recent theoretical predictions of Refs. \cite{Averin_94,Nazarov_Glazman,P_G}. Most surprisingly, we find that the conductance peak height exhibits a nonmonotonic dependence on temperature, which to our knowledge is experimentally observed here for the first time.

\begin{figure}[ht]
\includegraphics[width=1 \columnwidth]{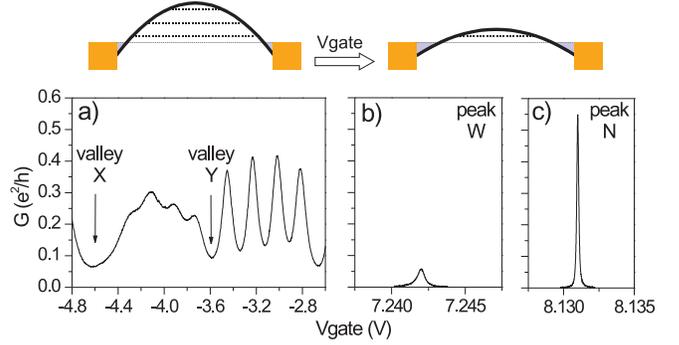}
\caption{\label{fig:overall}
(Color online) Tops schematic: the Schottky barriers (light gray triangles) become wider and less transparent as $V_{\rm gate}$ is applied. a) Differential conductance $G$ of the nanotube quantum dot measured \emph{vs.} $V_{\rm gate}$ for negative $V_{\rm gate}$ (large contact transparency). Two 4-electron ``shells'' are clearly visible. The ``X'' and ``Y'' marks indicate the wide valleys at which the data of Figure 2 are measured. $T=5$K. b,c) The same dot in the regime of small contact transparency; base temperature. The two peaks are further studied in Figures 3-5. }
\end{figure}

The nanotubes are grown on a Si/SiO$_2$ substrate by Chemical Vapor Deposition using CO as a feedstock gas \cite{Zheng2002}. This method usually produces single-wall nanotubes with diameters of about 2 nm. Individual nanotubes are contacted by long and narrow electrodes with a specific resistivity of $\sim 70 \Omega /\square$ and total resistance of several $k\Omega$. We present results measured on a single semiconducting nanotube with the two metal contacts separated by a distance of 400nm. All the measurements are performed in a dilution refrigerator at temperatures in the range of a few Kelvin to tens of mK.  

Figure 1 shows zero-bias conductance of the nanotube at representative ranges of positive and negative gate voltage, $V_{\rm gate}$. The metal electrodes form Schottky barriers to the valence band of the semiconducting nanotube, thereby defining a quantum dot. While the heights of the barriers are fixed, they get narrower and more transparent as the positive gate voltage moves the top of the valence band closer to the Fermi energy (schematic in Figure 1). Indeed, at positive $V_{\rm gate}$ corresponding to the wide Schottky barriers, the conductance demonstrates a set of very narrow peaks separated by wide Coulomb blockade valleys of vanishing conductance (Figure 1 b,c). For negative $V_{\rm gate}$ corresponding to the narrow Schottky barriers, the nanotube conductance shows wide single-electron peaks (Figure
1a). 

The peaks in Figure 1a form groups of four (``shells'') due to the orbital degeneracy in nanotubes \cite{4peaks}. For our initial measurements, we choose the conductance valleys at $V_{\rm gate} \approx -3.6$V and $-4.6$V, where a number of shells is completely filled. The large conductance background in these valleys is due to the elastic cotunneling processes \cite{Averin_Nazarov}. We find that this conductance, measured as a function of the bias voltage $V$, is flat up to at least $\sim 1$ meV at the temperatures of several K (not shown). Indeed, the excited states of the dot, corresponding to moving one electron to a higher shell, are separated by the energies of several meV from the ground state. Also, the ground state of the dot is non-degenerate, so that the Kondo effect \cite{GPCKreview} is not expected. Therefore, at the two selected $V_{\rm gate}$ we may consider the nanotube as a lumped tunneling junction of a known and constant transparency. 

As the temperature is lowered, in both valleys $G(V)$ gets suppressed, forming a narrow dip characteristic of environmental Coulomb blockade (Figure 2a). We replot the data for $V_{\rm gate} = -4.6$V in Figure 2b as $G(T,V)/G(T,0)$ \emph{vs.} $eV/k_B T$. (The logarithmic scale is chosen for the horizontal axis to present the data measured over several orders of magnitude in $V/T$. Also, the power laws with small exponent appear nearly linear on a log plot.) We find that the data measured at low bias and temperature collapse on the same curve, which is consistent with $G \propto \max(k_B T, eV)^{\alpha}$ at low $T$ and $V$. Moreover, the zero bias conductance at low temperatures is described very well as $G(T,0) \propto T^{\alpha}$ with $\alpha \approx 0.22$ in both valleys (Figure 2c). While we cannot measure the resistance of our leads directly, we have characterized the specific resistivity of the metal film and estimated the total resistance from the known lead geometry. The resulting $R_e \sim 3 k \Omega$ is consistent with $\alpha = 2 e^2 R_e/ h \approx 0.22$. This comparison should not be taken too literally, since the exponent is also affected by various shunting capacitances. We plug this exponent to the theoretical expression of Ref. \cite{Averin_Lukens} to generate the overlaying white line in Figure 2b, which accounts well for the scaling curve at different temperatures. 

\begin{figure}
\includegraphics[width=0.8\columnwidth]{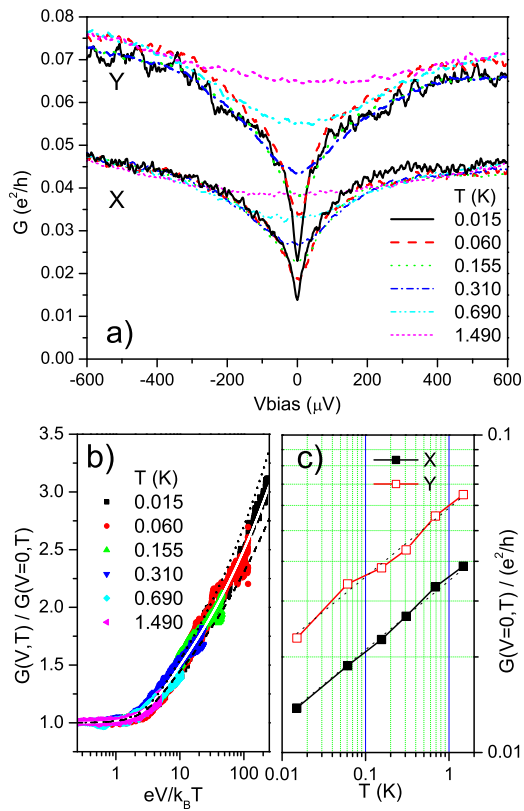}
\caption{\label{fig:ZBA}
(Color online) a) Differential conductance $G$ measured \emph{vs.} bias voltage $V$ in the middle of the two wide valleys marked by ``X'' and ``Y'' in Figure 1. b) $G(V,T)/G(V=0,T)$ plotted \emph{vs.} $eV/k_B T$ for valley X. Symbols of different colors and shapes correspond to the same set of temperatures as in panel (a). 
c) $G(V=0,T)$ plotted \emph{vs.} $T$ for the two valleys. We find that $G(V=0,T) \propto T^{\alpha}$ with $\alpha \approx 0.22$. This exponent is used to generate the overlaying white line in panel (b) according to the theoretical expression taken from Ref. \cite{Averin_Lukens}. The dashed and the dotted lines correspond to the same formula with $\alpha = 0.20$ and $\alpha = 0.24$ respectively. 
}
\end{figure}

Once the environment (\emph{i.e.} the leads) of the nanotube is characterized, we turn our attention to the range of large positive gate voltages, where the nanotube forms a well-isolated quantum dot. We focus on two single-electron conductance peaks in Figures 1b and c, centered at $V_{\rm gate} = -8.13$V (the ``narrow peak'', {\bf N}) and $V_{\rm gate} = -7.24$V (the ``wide peak'', {\bf W}). Figures 3a and b show the two peaks measured as a function of the gate voltage at zero bias in a range of temperatures. The dependences of their widths and height on temperature are summarized in Figure 4. Let us concentrate first on peak {\bf N}. At the highest temperatures, the peak shape is described by the standard expression for sequential tunneling through a single quantum level \cite{GlazmanMatveev,Beenakker}. (In practice, the peak shape predicted in Refs. \cite{GlazmanMatveev,Beenakker} is very close to the derivative of the Fermi function. In particular, the peak full width at half height is expected to be $\propto T$, which indeed is the case for the three highest temperatures in Figure 4a. At the lowest temperatures, the curve deviates from the linear dependence due to the contribution of the intrinsic (lifetime) broadening of the peak.

The height of peak {\bf N} increases as the temperature is lowered (Figure 4b). In the sequential tunneling through a single quantum level, the peak height should scale as $G_0 \propto \frac{1}{T} \frac{\Gamma_L \Gamma_R}{\Gamma_L+\Gamma_R}$, where $\Gamma_L$ and $\Gamma_R$ are the tunneling rates from the dot to the left and the right leads, respectively. For regular metallic leads, this results in $G_0 \propto 1/T$. Due to the environmental Coulomb blockade, electron tunneling rates $\Gamma_L, \Gamma_R$ to/from the quantum dot to the resistive leads should be suppressed at low temperatures and the $1/T$ dependence should be modified. A similar situation is encountered for a quantum dot coupled to Luttinger liquid leads, where the tunneling rates are also suppressed \cite{Auslaender,Postma}. In this case, it was shown theoretically that at low temperature the sequential tunneling expressions still hold as long as $\Gamma_{L,R} \lesssim T$ \cite{Furusaki}.

\begin{figure}
\includegraphics[width=0.8\columnwidth]{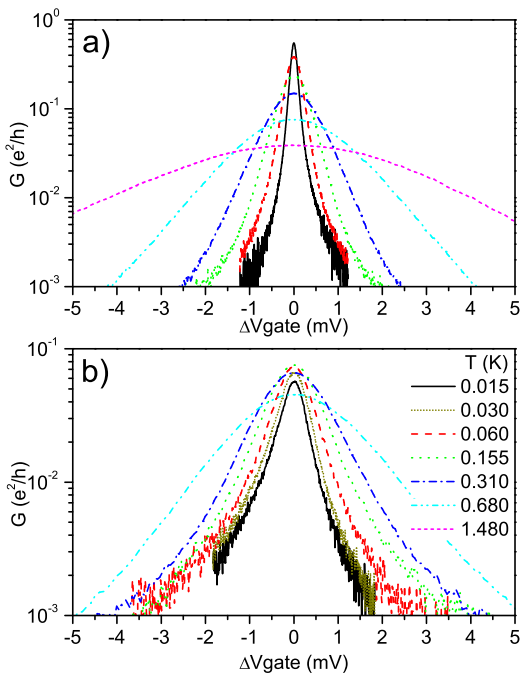}
\caption{\label{fig:peaks}
(Color online) Differential conductance for the peaks (a) {\bf N} and (b) {\bf W} vs. $\Delta V_{\rm gate}$, the gate voltage measured from their maxima. (The asymmetric tails of peak {\bf W} are due to spin degeneracy. The asymmetry may be lifted by magnetic field, Figure 5.)
}
\end{figure}

We expect that due to environmental blockade of tunneling $\Gamma_{L,R} \propto T^{\alpha_{L,R}}$ with $\alpha_{L,R} = 2 e^2 R_{L,R}/h$ \cite{NI}. Here $R_{L,R}$ are the resistances of the left and right leads, $R_L + R_R = R_e$, so that $\alpha_L + \alpha_R$ should be equal to $\alpha \approx 0.22$ which we extracted from Figure 2. Since the peak height even at the lowest temperatures is significantly less than $e^2/h$, we can assume that the tunneling rates between the dot and the left/right contacts are asymmetric: $\Gamma_L \gg \Gamma_R$. In this case, the expression for peak height reduces to $G_0 \propto \Gamma_R/T \propto T^{\alpha_R-1}$. Experimentally, we find that the height of peak {\bf N} at temperatures $T >$ 0.3K scales at $G_0 \propto T^{-0.85}$. Therefore, in our case $\alpha_R \approx 0.15$, which is slightly larger than $\alpha / 2 \approx 0.11$. The two leads to the nanotube were not designed to be identical in shape. Indeed, from the geometry of the leads we estimate that they are made of approximately 25 and 15 squares of the metal film. This difference may be partially responsible for $\alpha_R$ being larger than $\alpha / 2$. The value of $\alpha_{L,R}$ is also influenced by capacitances of the leads, which partially short out the lead resistances at the relevant frequency range. While in principle one can extract the exact exponents from the frequency-dependent impedance of the leads, the theory could not be easily applied to our case, because our metal leads have complicated shapes and non-uniform cross sections.

\begin{figure}
\includegraphics[width=0.8\columnwidth]{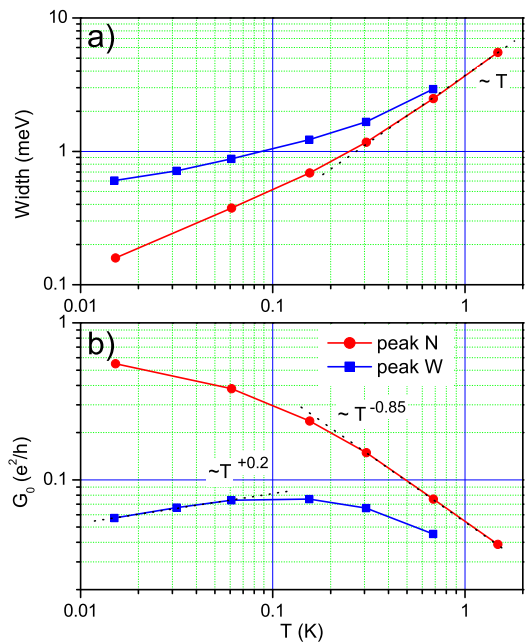}
\caption{\label{fig:heights}
(Color online) a) The width  and b) the maximal height of the peaks {\bf N} and {\bf W} measured as a function of temperature. The widths of the peaks are measured in the units of $V_{\rm gate}$. To convert these quantities to the actual energies, one has to multiply them by the gate efficiency factor \cite{QDreview}.  From the shape of the ``Coulomb diamonds'' in nonlinear conductance (not shown) we extract the gate efficiency factors of 0.10 and 0.11 for peaks {\bf N} and {\bf W} respectively. 
}
\end{figure}

For $\alpha_{L,R} <1$, the sequential tunneling picture should eventually break down as the temperature drops below $max(\Gamma_L, \Gamma_R)$. To study this resonant tunneling regime, let us turn our attention to the ``wide peak'', {\bf W} (Figure 3b). Surprisingly, as the width of this peak approaches saturation at the lower temperatures, the height of the peak starts to  \emph{decrease\/} (Figure 4b). 
We note that the saturation of the peak width  cannot be explained by the spurious effects of the external electromagnetic radiation, which could have saturated the effective electron temperature. Indeed, the width of peak {\bf N} keeps decreasing down to the lowest temperature. Other features do change as well, most importantly the height of peak {\bf W} and the depth of the zero-bias anomaly in Figure 2.

Recently, the resonant tunneling through a single quantum level with $\Gamma_{L,R} \propto T^{\alpha_{L,R}}$ has been considered theoretically \cite{Averin_94,Nazarov_Glazman,P_G}. Since the height of peak {\bf W} is always $<0.1 e^2/h$, the two tunneling rates are strongly asymmetric, $\Gamma_L \gg \Gamma_R$, in which case the width of the peak is predicted to saturate at the value of $\Gamma_0 \sim \Gamma_L(T_0)$. Here temperature $T_0$ is defined self-consistently as $k_B T_0 \sim \hbar \Gamma_L(T_0)$ \cite{Nazarov_Glazman,P_G}. As the temperature decreases below $T_0$, the peak conductance is predicted to decrease as $G_0 \propto T^{\alpha_L+\alpha_R}$. The exponent $\alpha_L+\alpha_R$ corresponds to a direct tunneling from one lead to the other through the wide resonance level ($\Gamma_0 > T$). Indeed, we find that the height of peak {\bf W} at lowest temperatures decreases approximately as $\propto T^{0.2}$. This exponent is very close to $\alpha  \approx 0.22$ we have extracted from Figure 2. The small difference is likely caused by not following the decrease of the peak over a sufficient range of temperatures.

\begin{figure}
\includegraphics[width=0.8\columnwidth]{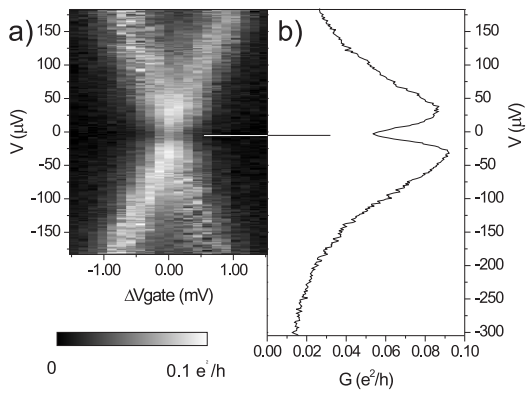}
\caption{\label{fig:map}
a) The differential conductance map of peak {\bf W} measured as a function of the gate voltage and source-drain bias at the base temperature. The X-shaped feature corresponds to the alignment of the resonance level with the chemical potential of the source or the drain. The cross-section of the conductance map along the horizontal line $V=0$ corresponds to the peak shape as measured in Figure 3a. b) Conductance as a function of the source-drain bias measured at the gate voltage corresponding to the center of the peak. In both (a) and (b), the sample is subject to perpendicular magnetic field of 4T applied to split the spin degeneracy. 
}
\end{figure}

To get a better insight into the temperature evolution of conductance in the resonant tunneling regime, we present a two-dimensional map of the wide peak conductance measured as a function of both $V_{\rm gate}$ and $V$ (Figure 5a). Figure 5b shows the vertical cross section of the conductance map measured as a function of the bias voltage $V$ at the center of the peak, $\Delta V_{\rm gate}= 0$ . The most remarkable feature of Figure 5 is the zero-bias suppression, visible both in the map and the cross section. This is the same dissipation-induced feature as observed in Figure 2. It is now superimposed on top of the wide resonance level ($\Gamma_0 \sim 100 \mu V$) \cite{P_G}. It appears quite natural that as the temperature is raised, the zero-bias suppression will be washed away first, resulting in the increase of the zero-bias conductance. On the other hand, the width of the peak will not be immediately affected until $T$ becomes of the order of $\Gamma_0$. This exactly corresponds to our findings in Figure 3b. 

In summary, we study the suppression of conductance through a nanotube connected to resistive leads, which create a dissipative environment for the tunneling electrons. We focus our attention on the transition from the sequential to the resonant tunneling through a single level in the nanotube. For the fist time, we observe a nonmonotonic temperature dependence of the conductance peak height. In the resonant tunneling regime, as the temperature decreases, the width of the conductance peak saturates, but its height starts to \emph{decrease}. In this regime, one should view the transport as a coherent tunneling of electrons between the two leads mediated by the resonant level. In agreement with the existing theory, this process is suppressed stronger than individual tunneling rates between each of the leads and the dot.

Acknowledgements: We appreciate valuable discussions with D. Averin, H. Baranger, M. Gershenson, L. Glazman, I. Gornyi, J. Liu, C. Marcus, K. Matveev and D. Polyakov. The work was supported by NSF DMR-0239748.


\begin{thebibliography}{99}
\bibitem{note} Unlike the ``conventional'' Coulomb blockade \cite{QDreview}, the environmental Coulomb blockade does not require an isolated nanoscale island and is observed in a single tunneling junction.
\end{thebibliography}
\end{document}